\documentclass[review]{elsarticle}
\usepackage{lineno,hyperref}
\modulolinenumbers[5]

\journal{Journal of Atmospheric and Solar-Terrestrial Physics}

\biboptions{authoryear}

\begin{document}

\begin{frontmatter}

\title{Modulation of Solar Wind Energy Flux Input on Global Tropical Cyclone Activity}


\author[a]{Hui Li\corref{mycorrespondingauthor}}
\cortext[mycorrespondingauthor]{Corresponding author}
\ead{hli@nssc.ac.cn}

\author[a,f]{Chi Wang}
\author[b]{Shengping He}
\author[c,b,e]{Huijun Wang}
\author[d]{Cui Tu}
\author[a]{Jiyao Xu}
\author[e]{Fei Li}
\author[a]{Xiaocheng Guo}

\address[a]{State Key Laboratory of Space Weather, National Space Science Center, CAS, Beijing, 100190, China}
\address[b]{Climate Change Research Center, CAS, Beijing, 100029, China.}
\address[c]{Nanjing University for Information Science and Technology, Collaborative Innovation Center on Forecast and Evaluation of Meteorological Disasters/Key Laboratory of Meteorological Disaster, Ministry of Education, Nanjing, 210044, China.}
\address[d]{Laboratory of Near Space Environment, National Space Science Center, CAS, Beijing, 100190, China.}
\address[e]{Nansen-Zhu International Research Center, Institute of Atmospheric Physics, CAS, Beijing, 100029, China.}
\address[f]{University of Chinese Academy of Sciences, Beijing, 100049, China.}

\begin{abstract}
Studies on Sun-climate connection have been carried out for several decades, and almost all of them focused on the effects of solar total irradiation energy. As the second major terrestrial energy source from outer space, the solar wind energy flux exhibits more significant long-term variations. However, its link to the global climate change is rarely concerned and remain a mystery. As a fundamental and important aspect of the Earth's weather and climate system, tropical cyclone activity has been causing more and more attentions. Here we investigate the possible modulation of the total energy flux input from the solar wind into the Earth's magnetosphere on the global tropical cyclone activity during 1963--2012. From a global perspective, the accumulated cyclone energy increases gradually since 1963 and start to decrease after 1994. Compare to the previously frequently used parameters, e,g., the sunspot number, the total solar irradiation, the solar F10.7 irradiation, the tropical sea surface temperature, and the south oscillation index, the total solar wind energy flux input exhibits a better correlation with the global tropical cyclone activity. Furthermore, the tropical cyclones seem to be more intense with higher geomagnetic activities. A plausible modulation mechanism is thus proposed to link the terrestrial weather phenomenon to the seemly-unrelated solar wind energy input.
\end{abstract}


\end{frontmatter}


\section{Introduction}
\label{sec:intro}

Since 1970, the affecting factors of global climate change have gradually becoming the central topics in the science community, the general public, and the policy-makers as well. Besides human activities, solar variability is another significant natural contributor. \citet{Herschel 1801} first proposed that solar activity may affect the Earth's climate system. Thereafter, various studies confirmed that the Earth's weather and climate changes are significantly affected by solar activity variations. The time scale ranges from several hours or several days \citep{Tinsley and Deen 1991,Tinsley and Heelis 1993,Veretenenko and Thejll 2004,Kniveton and Tinsley 2004} to a decade or a century, or even longer time scale \citep{Friis-Christensen and Lassen 1991,Bond et al 2001,Neff et al 2001,Wang et al 2005,Camp and Tung 2007,van Loon et al 2007}.

There are two major forms of the solar irradiance, electromagnetic radiation and corpuscular radiation. The electromagnetic radiation is so-called solar photons, and the corpuscular radiation is referred to the solar wind energy flux here, which contains the solar wind, the interplanetary magnetic field, and solar energetic particles. In general, the energy content of electromagnetic radiation enters into the terrestrial system is of 4-5 orders higher than the solar wind energy flux, therefore it is no surprise that most of the researches about the solar influences on the Earth's climate change focused on the effects of solar electromagnetic radiation. However, the variation amplitude of electromagnetic radiation during a solar cycle (about 11 years) is quite small, of only about 1 Wm$^{-2}$ or $\sim$ 0.1\% \citep{Frohlich 2013}. In contrast, the decadal variation of the solar wind energy flux could even exceed 100\%, which makes the absolute variation of solar wind energy flux comparative to solar electromagnetic radiation. Massive solar wind energy flux entering into geospace ($E_{in}$) via magnetic reconnection or viscous interaction can heat the Earth's atmosphere by two major aprroaches, auroral particle precipitation and Joule heating \citep{Vasyliunas 2011}, and it may drive the Earth's weather and climate change through some nonlinear interaction mechanisms and enlarge its effects significantly.

As a fundamental and important aspect of the Earth's weather and climate system, tropical cyclone (TC) is cyclonic circulations typically forming over the tropical and subtropical oceans. On the average, there are about 80 TCs throughout the world every year. It is estimated that the power of a mature TC is as high as 6$\times$10$^{14}$ W. In general, TCs can transport the thermal energy from the equator to high-latitudes and contribute to maintain the global heat balance. In the meantime, TC is one of the diaster weather phenomena, which could cause very enormous loss of life and property. \citet{Nicholls et al 1995} stated that more than 1.9 million deaths are associated with severe TCs during 1780-1970. Thus, TCs have been causing more and more attentions since the 21st century \citep{Webster et al 2005, Emanuel 2005}. \citet{Webster et al 2005} have shown that the TC number and cyclone days decreased in all cyclone basins except for the North Atlantic during 1995-2005, however, the number and proportion of Category 4-5 hurricanes has a large enhancement. 

So far, it is believed that the intensity of a TC or the active level of TCs during a season cannot be attributed to a single factor, such as the global warming or other environment change. \citet{Cohen and Sweetser 1975} suggested the correlation between solar cycle and Atlantic TC activities from the similarities in the spectra for the 7-yr running mean TC number in North Atlantic, the 7-yr running mean length of the cyclone season, and the 12-month running mean sunspot numbers. \citet{Ivanov 2007} later confirmed the correlation between magnetic storms and TCs in the Atlantic, and found that the linear correlation coefficient changed in different regions from positive to negative values. \citet{Elsner and Jagger 2008} also reported that the relationship between the number of sunspots and hurricane activity over the Caribbean was negative. \citet{Elsner et al 2010} later found that changes in solar ultraviolet radiation (UV) are the major cause. They believed that TCs can reinforce the effect of relatively small changes in solar UV output and thereby fairly influence the Earth's climate through the TCs energy dissipation by ocean mixing and atmospheric transport. Recently, \citet{Ge et al 2015} confirmed the high sensitivity of the TC warm core to solar shortwave radiative effect; \citet{Haig and Nott 2016} showed that solar forcing (the number of sunspots) contributes to the TC activities over decadal, interdecadal, and centennial scales.

Most previous studies on the relationship between solar activity and TCs focused on the effects from solar electromagnetic radiations. To deepen the understanding of TC activities variation tendency and to improve the prediction accuracy of the climate model, it is worthy to explore other possible driving factors. In this study, we pay our attention on the influence of solar wind energy flux on TCs activities. This paper is organized as follows: the data sets are described in section 2; the results are given in section 3; and the plausible mechanism and summary are presented in section 4 and 5.

\section{Data Sets}
\label{sec:meth}

\subsection{TC activities}

International Best Track Archive for Climate Stewardship (IBTrACS) project aims at merging tropical storm information from all the regional specialized meteorological centers and other international centers and individuals into one product, and providing best track data of TCs in a centralized location. The IBTrACS project checks the quality of storm inventories, positions, wind speeds, and pressures. It contains the most complete global set of historical TCs, and is endorsed to be an official archiving and distribution resource for TC best track data by the World Meteorological Organization (Tropical Cyclone Programme). In this study, global TC activity was tabulated by using the IBTrACS Dataset v03r05 from 1963 to 2012 for all tropical cyclone basins. During this half century, a total of 6238 TC events have been recorded in the global context.

ACE for each TC is defined as the sum of the square of 1-minute surface wind speed maximum at 6-hour intervals during the cyclone lifetime \citep{Bell et al 2000}. Annual ACE is the sum of the ACEs for each cyclone in the year. It takes into consideration the number, intensity, and duration period of all the TCs in a year, and can represent the kinetic energy generated by TCs.

\subsection{Solar wind energy input}
It is still a great observational challenge to accurately monitor the solar wind energy input into the Earth's magnetosphere on a global scale. Nevertherless, the global three-dimensional magnetohydrodynamic simulation (3D MHD) model makes it possible to do some estimations. Our previous work \citep{Wang et al 2014} performed 3D global MHD simulations and proposed a empirical formula to estimate the solar wind energy flux input, which is given as follows:
\begin{linenomath*}
\begin{equation}
E_{in} \mathbf{(W)} = 3.78\times10^7\times n_{SW}^{0.24}\times V_{SW}^{1.47}\times B_T^{0.86}\times \left[\sin^{2.70}\left(\frac{\theta}{2}\right)+0.25\right]
\end{equation}
\end{linenomath*}
Here, $E_{in}$ represents the solar wind energy flux into the magnetosphere in the unit of watts. $n_{SW}$, $V_{SW}$, and $B_T$ ($=\sqrt{B_Y^2+B_Z^2}$) is the solar wind number density in the unit of cm$^{-3}$, the solar wind velocity in the unit of km/s, and the transverse magnetic field magnitude in the unit of nT, respectively. $\theta$ is the interplanetary magnetic field clock angle. 

The solar wind parameters can be obtained from the OMNI project, which primarily makes a compilation of hourly-averaged solar wind magnetic field and plasma parameters from several spacecrafts since 1963. All the spacecrafts are in geocentric or L1 (Lagrange point) orbits. The data have been extensively cross compared or cross-normalized, and are well used in space physics studies. Based on the above energy coupling function, the solar wind energy flux entering into the magnetosphere can be obtained when the OMNI 2 data sets are available.

\subsection{Geomagnetic activities}
In this study, the relative ap$_{max}$ is defined to represents the level of geomangeitc activities during TCs, which is obtained as follows
\begin{linenomath*}
\begin{equation}
\label{wr4}
   \mathbf{relative\ \ \ } ap_{max} = ap_{max}/\overline{ap_{max}}
\end{equation}
\end{linenomath*}
where ap$_{max}$ is the maximum ap index during a TC event. $\overline{ap_{max}}$ is calculated based on Monte Carlo method. The time duration of a concerned TC event is recorded as $\Delta$T. We randomly choose a begin time T$_0$ from 1964 to 2012, and then obtain the maximum ap index from T$_0$ to T$_0$+$\Delta$T. Then, we repeat the above steps 10$^6$ times. $\overline{ap_{max}}$ is the mean value of the 10$^6$ maximum ap index. When the relative ap$_{max}$ is much greater than 1, it represents that the TC event is during a very disturbed geomagnetic environment. When the relative ap$_{max}$ is much less than 1, it represents that the TC event is during a very quiet geomagnetic period.

\section{Results}
\label{sec:res}

\begin{figure}[htbp]
\centering
\noindent\includegraphics[width=13pc]{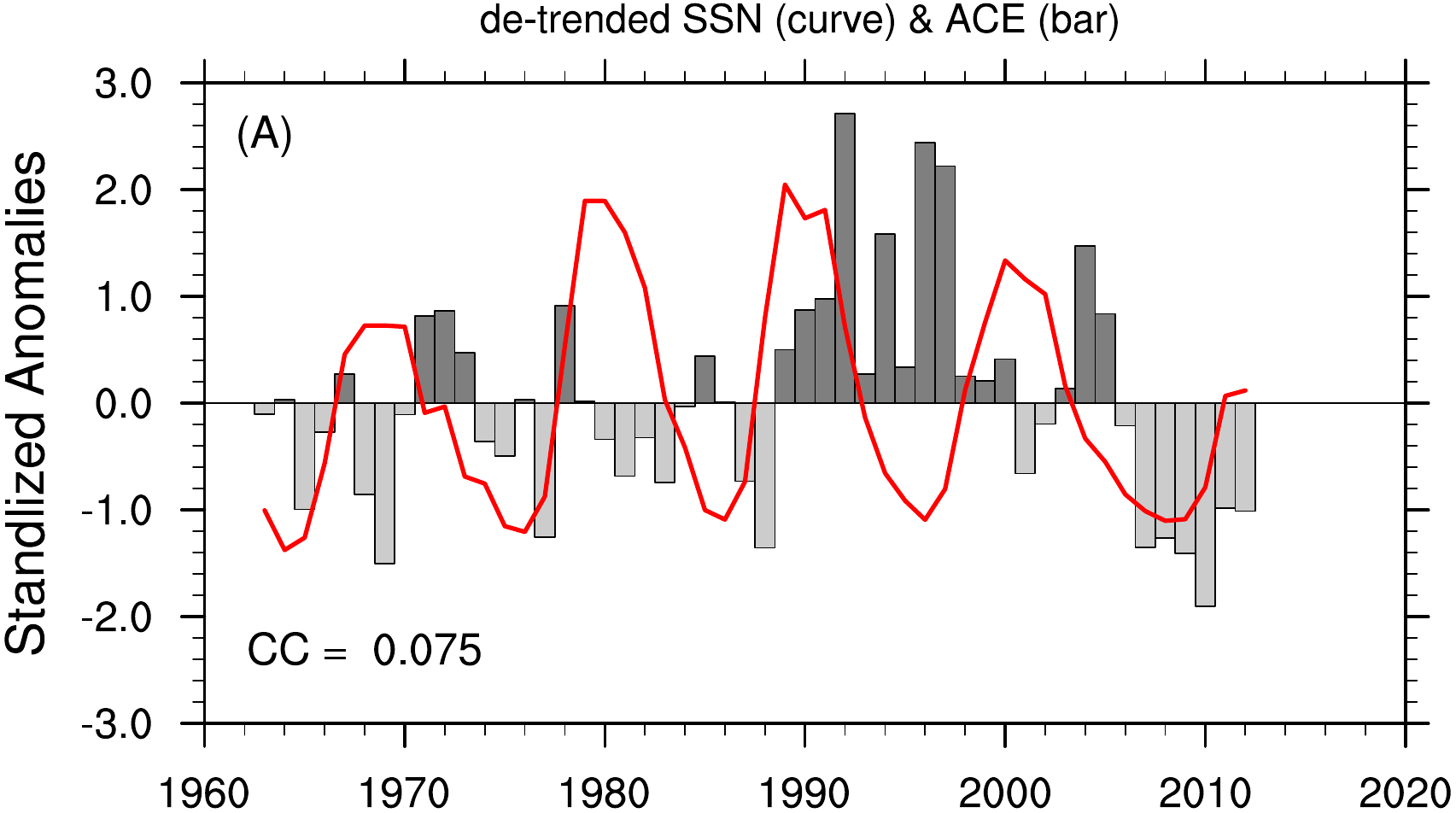}
\noindent\includegraphics[width=13pc]{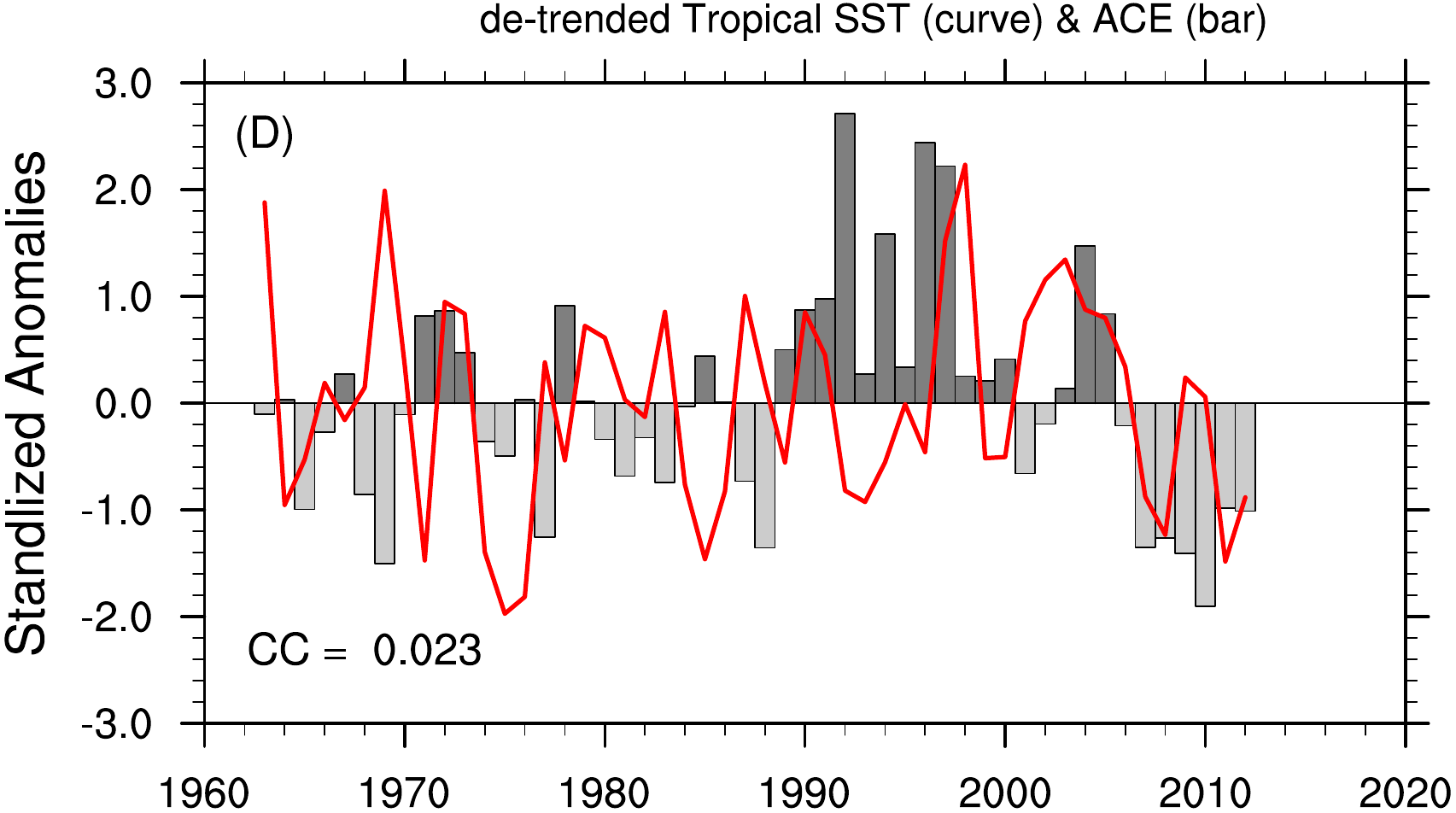}
\noindent\includegraphics[width=13pc]{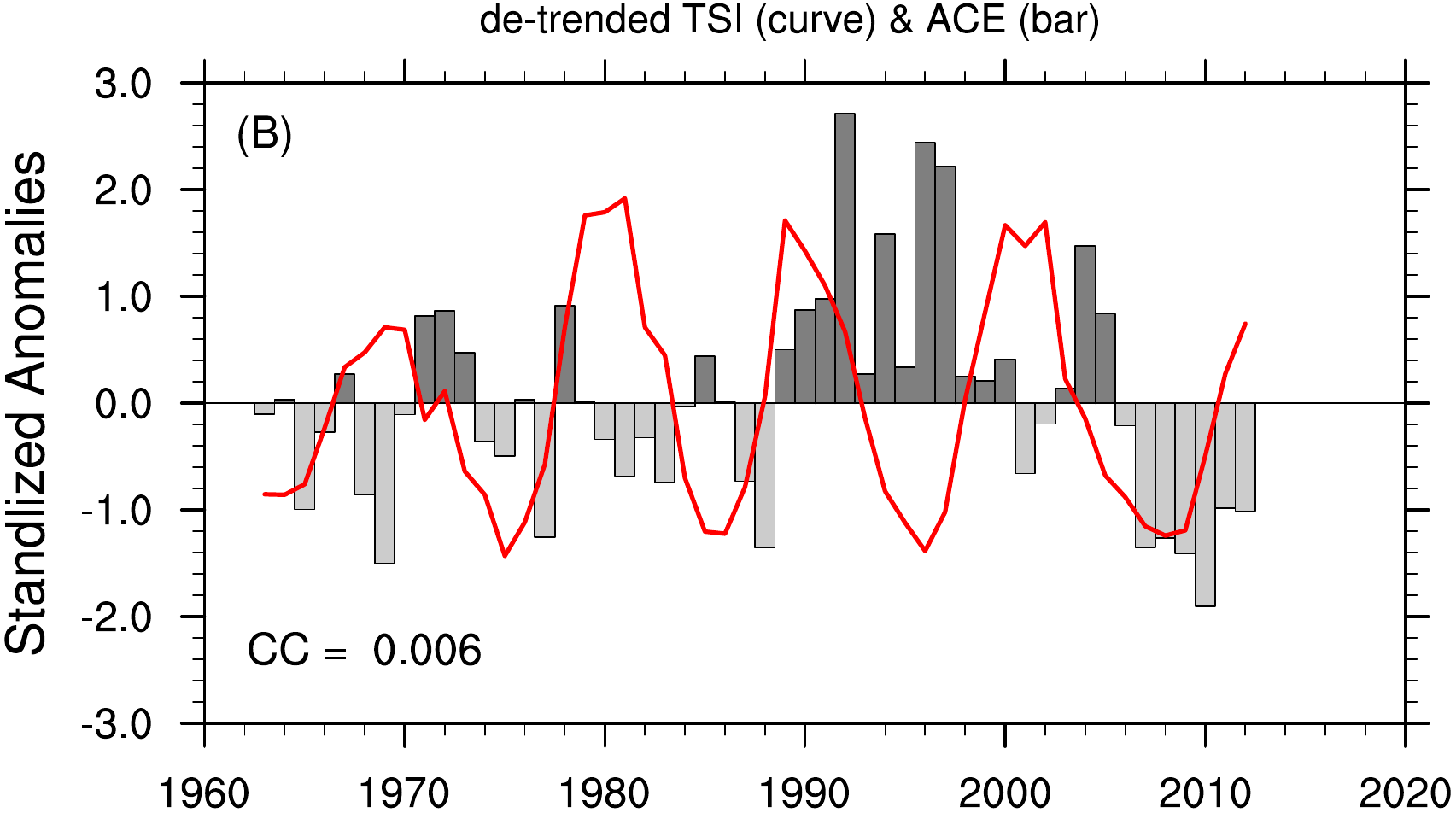}
\noindent\includegraphics[width=13pc]{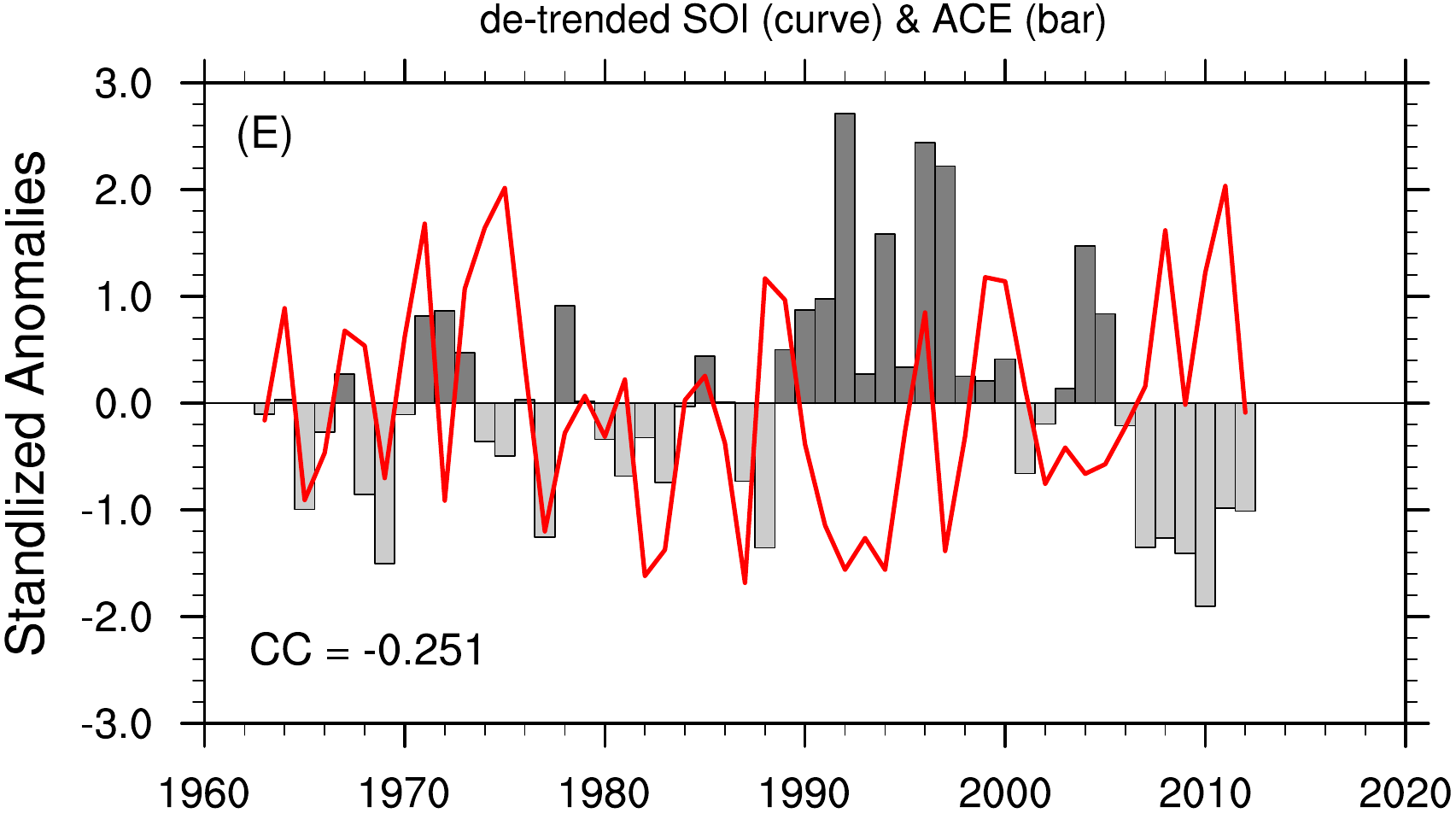}
\noindent\includegraphics[width=13pc]{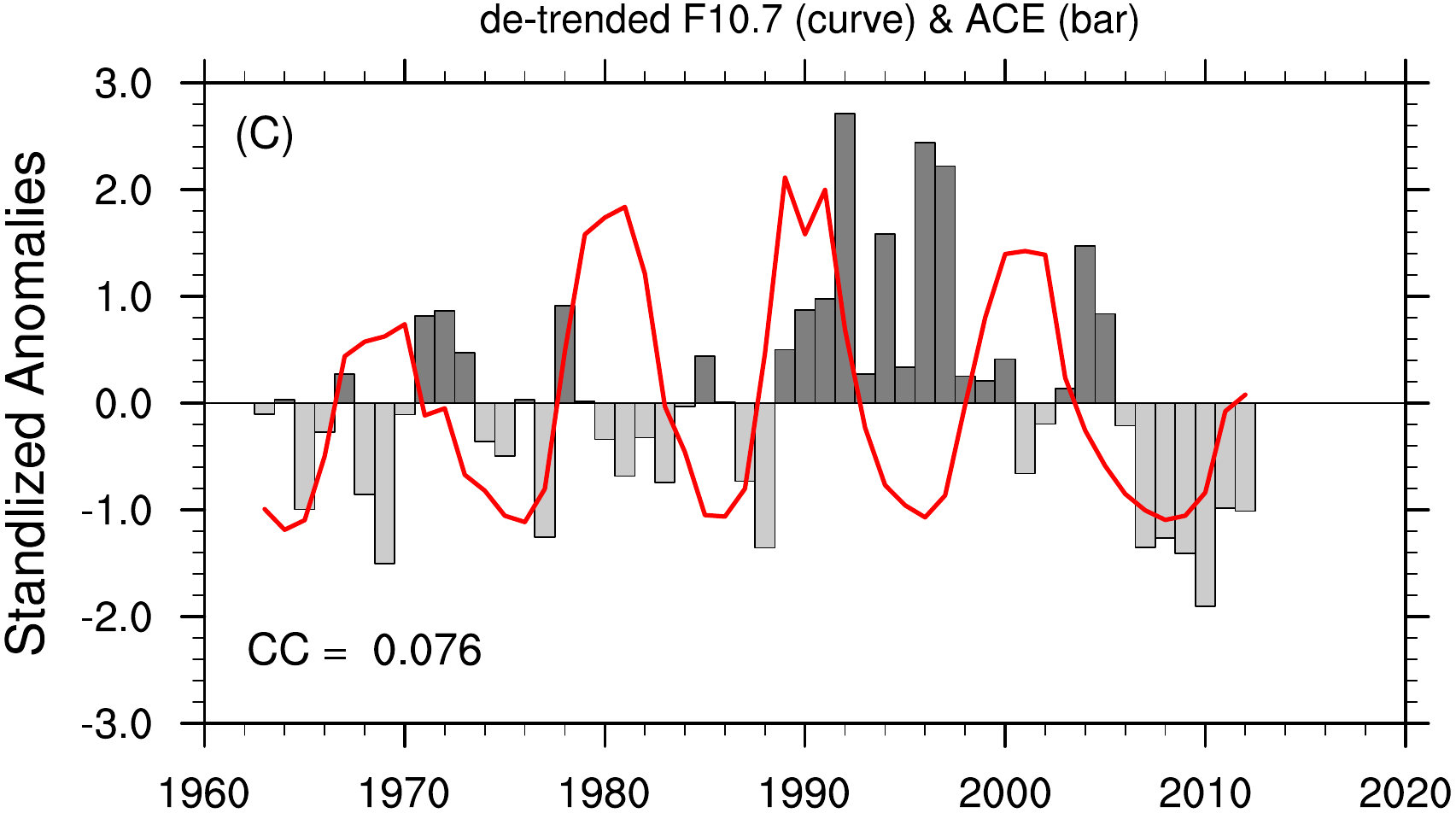}
\noindent\includegraphics[width=13pc]{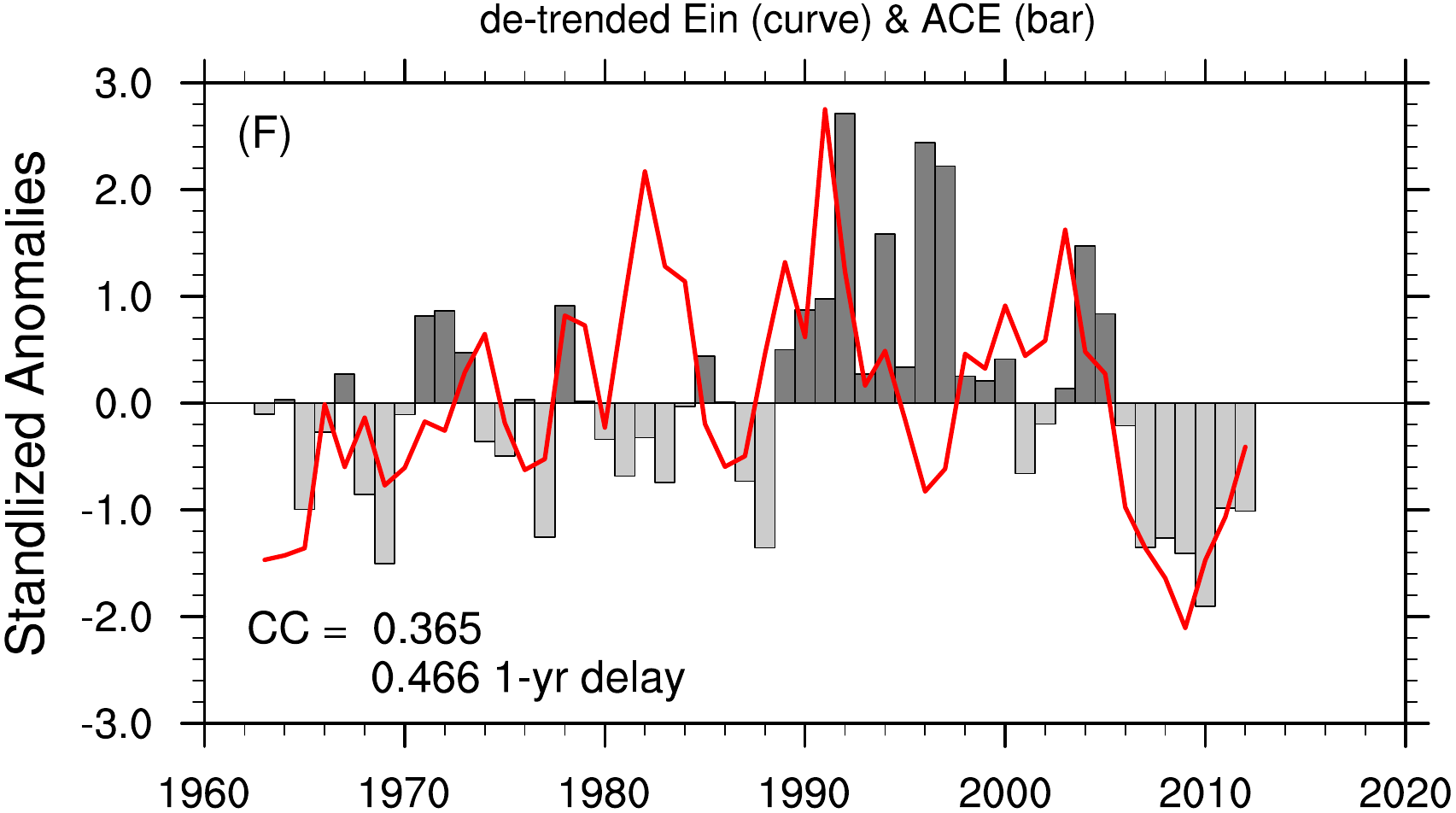}
\caption{Overview of the relationships of the annual parameters, e.g., SSN (A), TSI (B), F107 (C), SST (D), SOI (E), and $E_{in}$ (F) on the global tropical cyclone activity intensity, the annual ACE over all TC basins from 1963 to 2012. Note that the detrending and normalizing processes are all made for these parameters. The threshold value is 0.235, 0.279, and 0.361 for correlation at the 90\%, 95\%, and 99\% confidence level, respectively.}
\label{overview}
\end{figure}

We present the relationships of the annual parameters, e.g., sunspot number (SSN, Fig. 1A), total solar irradiation (TSI, Fig. 1B), solar F10.7 irradiation (F107, Fig. 1C), tropical sea surface temperature (SST, Fig. 1D), south oscillation index (SOI, Fig. 1E), the total energy flux input parameter ($E_{in}$, Fig. 1F) with the global tropical cyclone activity intensity indicated by the annual ACE over all TC basins from 1963 to 2012. Note that, the detrending and normalizing processes are made for these parameters. SSN, TSI and F107 all show obvious 11-year periodic variations because of the variability of solar activity. However, their long-term variations are not significant. SST represents a lasting gradual enhancement until 2005, thereafter, SST remains at that level. SOI has a 4-year or 5-year periodic variation. During 1975-2007, the SOI is almost negative, indicating an El Nino episodes in the Pacific Ocean. $E_{in}$ represents a clear 11-year variation as well, with the peak value in 1991 and the minimum value in 2009. For the long-term variation, $E_{in}$ represent a significant enhancement before 1987, and then a gradual decreases till now. ACE has a peak around 1992. During the latter 20 years, only the decreasing of long-term variation of $E_{in}$ matches accordingly the lasting decrease of ACE. The correlation coefficient between $E_{in}$ and ACE is 0.365, which is much stronger than the others, e.g., 0.075 for SSN, 0.006 for TSI, 0076 for F107, 0.023 for SST, -0.251 for SOI. The threshold value is 0.235, 0.279, and 0.361 for correlation at the 90\%, 95\%, and 99\% confidence level, respectively. It thus concludes that the TC activity (represented by annual ACE) is only correlated with $E_{in}$, but not with the SSN, TSI, F107, SST and SOI.

The ring current energy content is one of the two major energy sinks of solar wind energy flux input into the magnetosphere. The main carriers of the storm ring current are protons with energy from several keV to a few hundred keV \citep{Daglis et al 1999}. There are dominant loss processes of the ring current energetic ions. One is the Coulomb collision with cold dense plasmas in the plasmasphere \citep{Wentworth et al 1959}. The other one is the charge exchange with neutral atoms \citep{Dessler and Parker 1959}.Recently, \citet{Ebihara et al 2014} evaluated the Coulomb lifetime and Change exchange lifetime of ring current ions. Based on their results, the Coulomb lifetime and Change exchange lifetime are estimated to be $\sim$ 330 days and $\sim$ 220 days for the ring current protons with the energy of 200 keV at L = 4.5. Thus, 1-yr time delay of the modulation of $E_{in}$ on TC activities is expected. The correlation coefficient between $E_{in}$ and ACE is indeed even better of 0.466 if 1-yr time delay is considered.

\begin{figure}[htbp]
\centering
\noindent\includegraphics[width=13pc]{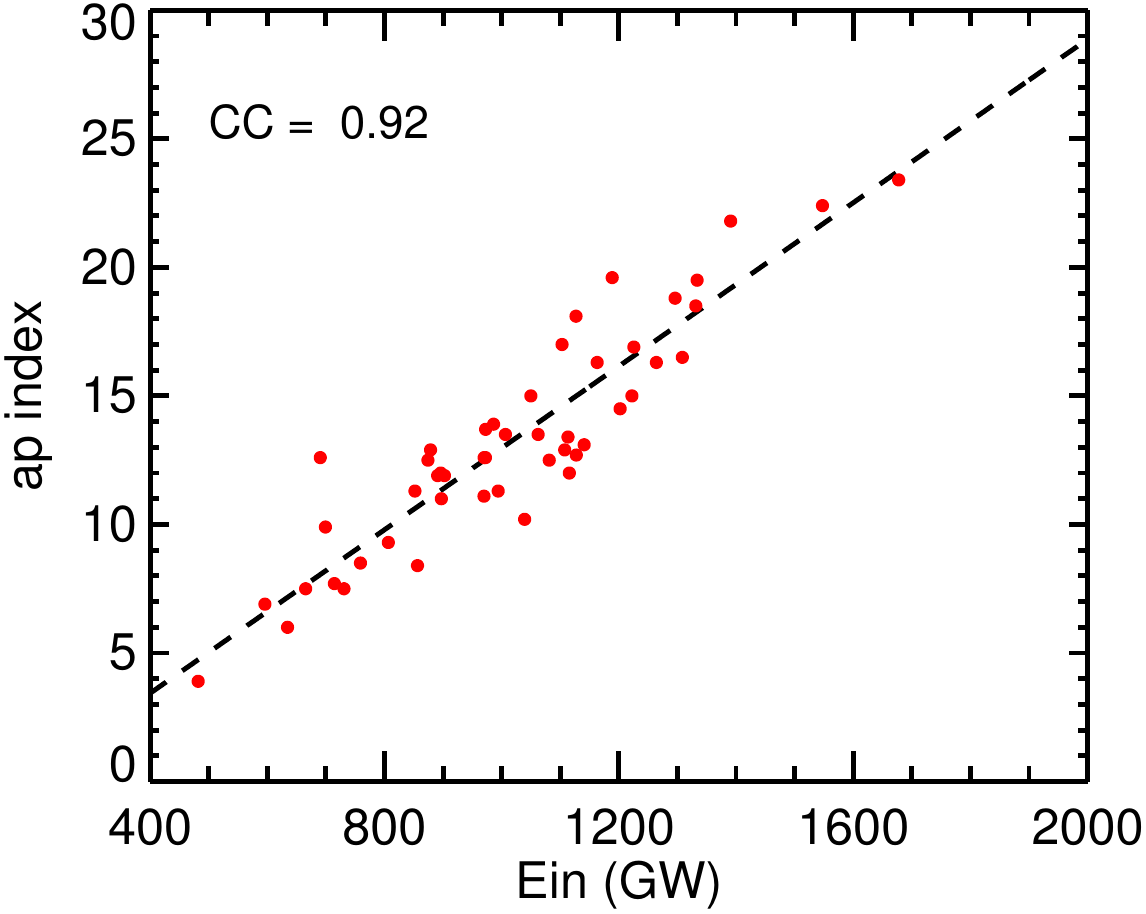}
\noindent\includegraphics[width=13pc]{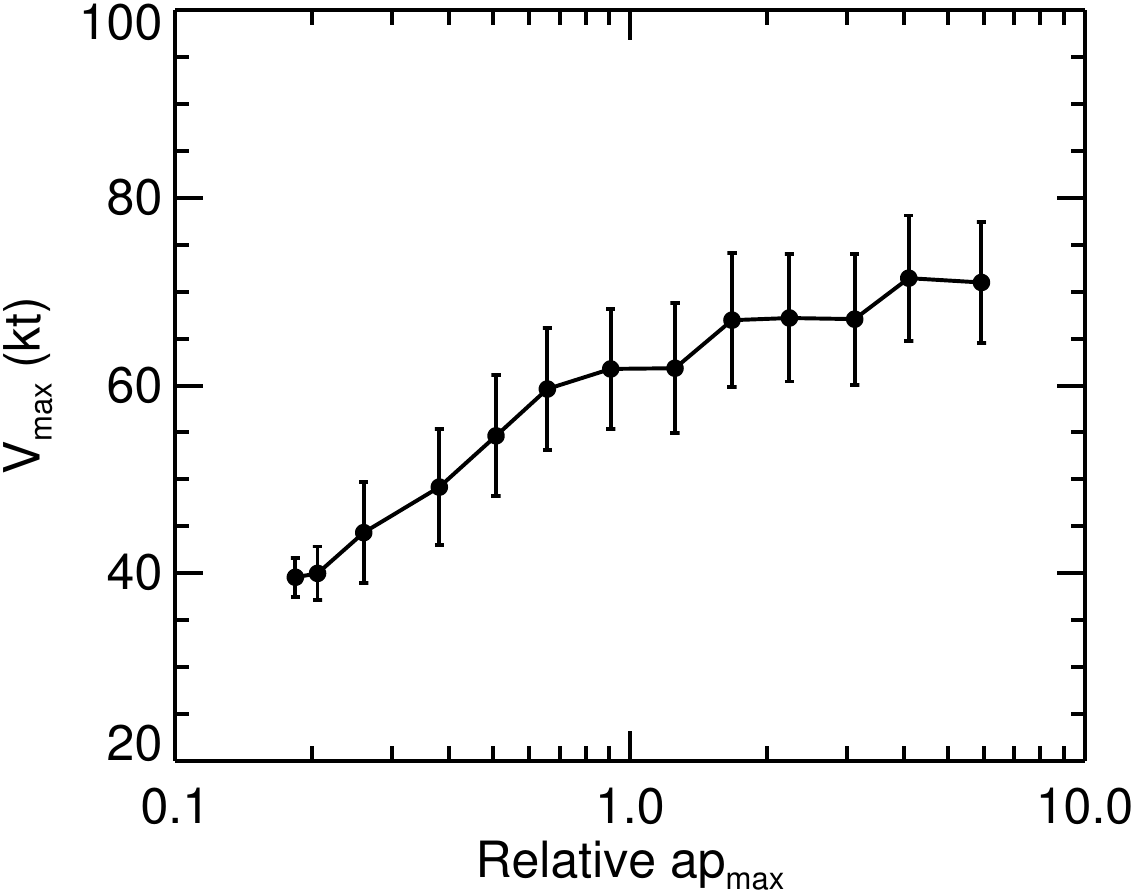}
\caption{Relationship between the solar wind energy flux input and TC intensity and geomagnetic activity intensity. (A) linear relationship between the solar wind energy input and the ap index, with the correlation coefficient of 0.92; (B) the maximum wind speed of TC events during different relative relative ap$_{max}$.}
\label{ap}
\end{figure}

Traditionally, areas of tropical cyclone formation could be divided into six basins, including the western Pacific Ocean (WP), the eastern Pacific Ocean (EP), the southern Pacific Ocean (SP), the northern Atlantic Ocean (NA), the northern Indian Ocean (NI), and the southern Indian Ocean (SI). We further study the relationship between the solar wind energy flux input and regional tropical cyclone activities. Tropical cyclone activities reveal a positive correlation relationship with $E_{in}$ at the 95\% confidence level at the basin of WP, EP, and SP (0.414, 0.379, 0.315), a negative correlation of -0.285 at the basin of NA, and no close relationship at the basin of NI and SI.

If the solar wind energy flux can indeed modulate the global TC activity as shown before, a natural thought is that a TC activity should be more intense during the period when more solar wind energy flux enters into the magnetosphere. In the early age of space era, there are many data gaps of $E_{in}$ for many TCs. Magnetospheric studies reveals that the more solar wind energy flux input would cause more intense geomagnetic disturbances \citep{Li et al 2012}. The well-used geomagnetic index, 3-hourly ap index, has no data gap for any TC and thus is used as a proxy of $E_{in}$ used here. As shown in Fig. 2A, there is a linear relationship between the solar wind energy input and the ap index, with the correlation coefficient of 0.92. The mentioned natural thought is confirmed by Fig. 2B, which clearly shows  that the maximum wind speed of TC events during severe geomagnetic activities tends to be greater than that during quiet geomagnetic activities.

\section{Plausible Mechanism}
\label{sec:disc}

The modulation mechanism of solar wind energy flux on TC activity remains mysterious. However, previous studies might give some clues. The solar wind energy flux is the primary energy source for the magnetosphere. \citet{Vasyliunas 2011} summarized the energy conversion and dissipation/loss processes in the magnetosphere. In short, the solar wind energy flux can heat the atmosphere unevenly by ring current precipitation, auroral electron precipitation, and Joule heating, especially during geomagnetic active time period. Such atmosphere heating is usually occurred at the altitude of 80-200 km, which is called thermosphere. The detailed coupling between thermosphere and troposphere is not clear so far. However, the modulation of thermospheric temperature by solar wind energy flux could cause indirectly dynamic variation in the lower atmosphere \citep{Kodera and Kuroda 2002}.

\begin{figure}[htbp]
\centering
\noindent\includegraphics[width=22pc]{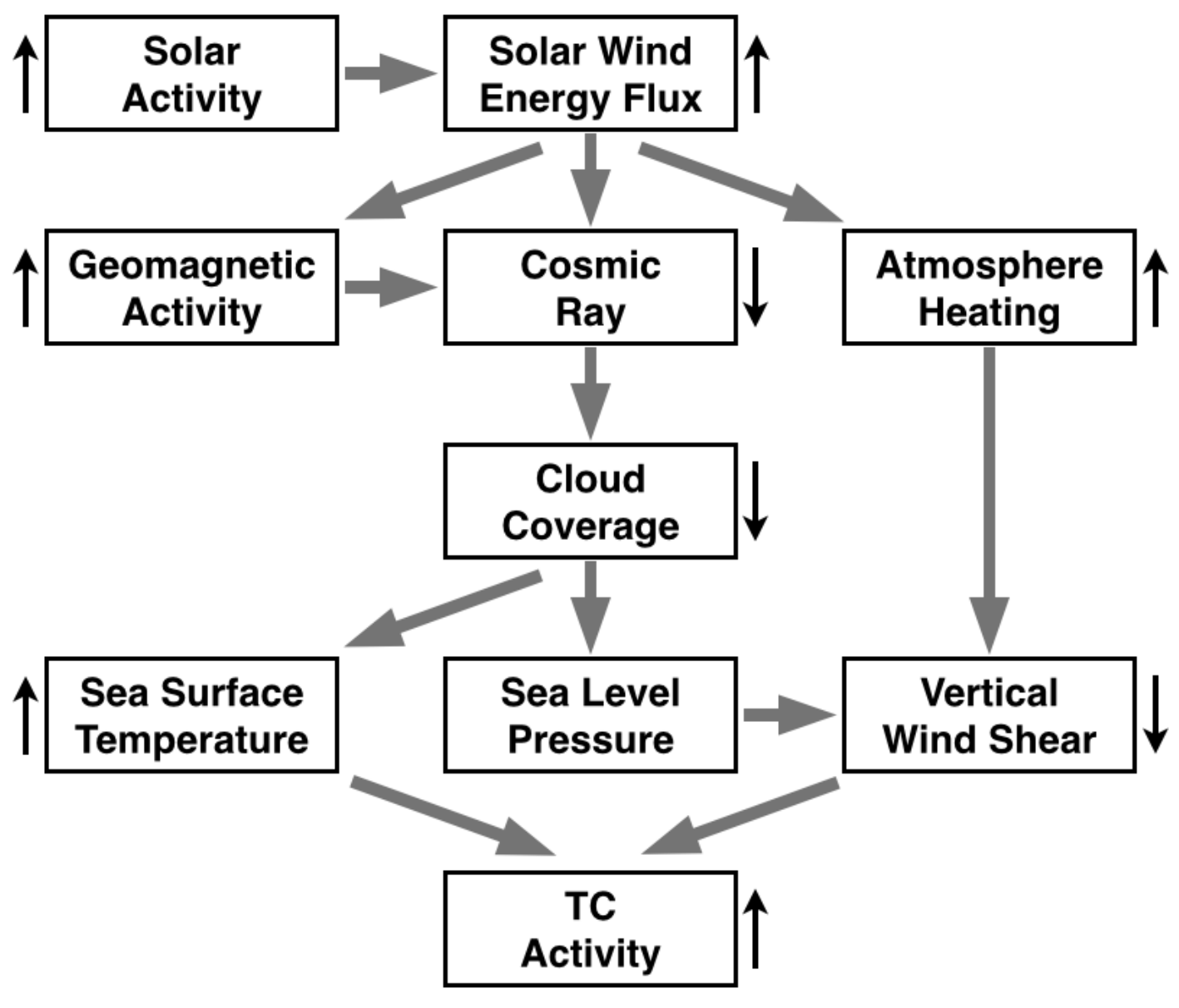}
\caption{Sketch of modulation mechanism.}
\label{sketch}
\end{figure}

We proposed a plausible mechanism to interpret the possible modulation process, as shown in Fig. 3. As solar activities enhance, the solar wind energy flux increases, resulting in 1) geomagnetic activities enhancements \citep{Li et al 2012}, 2) reduction of cosmic rays reaching the atmosphere due to an enhanced IMF shielding \citep{Singh and Singh 2008}, and 3) enhancement of atmosphere heating \citep{Vasyliunas 2011}. Under an enhanced geomagnetic field environment, the transport coefficient of cosmic rays along the mean magnetic field decreases \citep{Giacalone and Jokipii 1999}, causing less cosmic rays could reach the atmosphere as well. As a verification, the correlation coefficient between the annual solar wind energy flux and the cosmic ray intensity at Oulu station is -0.79 in our study. The global cloud coverage was found to be positively correlated with cosmic ray flux \citep{Svensmark and Friis-Christensen 1997}. Thus, the global cloud coverage (tcdc) decreases for enhanced solar activities,
and  the sea water would absorb more solar irradiance energy and result in an increase of the sea surface temperature (SST) and latent heat. As a major energy source, latent heat release could issue in rapid intensification and development of tropical cyclones \citep{Pauley and Smith 1988, Kuo and Low-Nam 1990}. Meanwhile, the gradient of sea level pressure (SLP) from the continent to the sea enhances as the tcdc decrease. The enhancements of the SLP gradient and the atmosphere heating contribute to the reduction of the vertical wind shear over the tropical oceans, which leads to an enhancement of TC activity \citep{Gray 1968, Gray et al 1993}.

\begin{figure}[htbp]
\centering
\noindent\includegraphics[width=13pc]{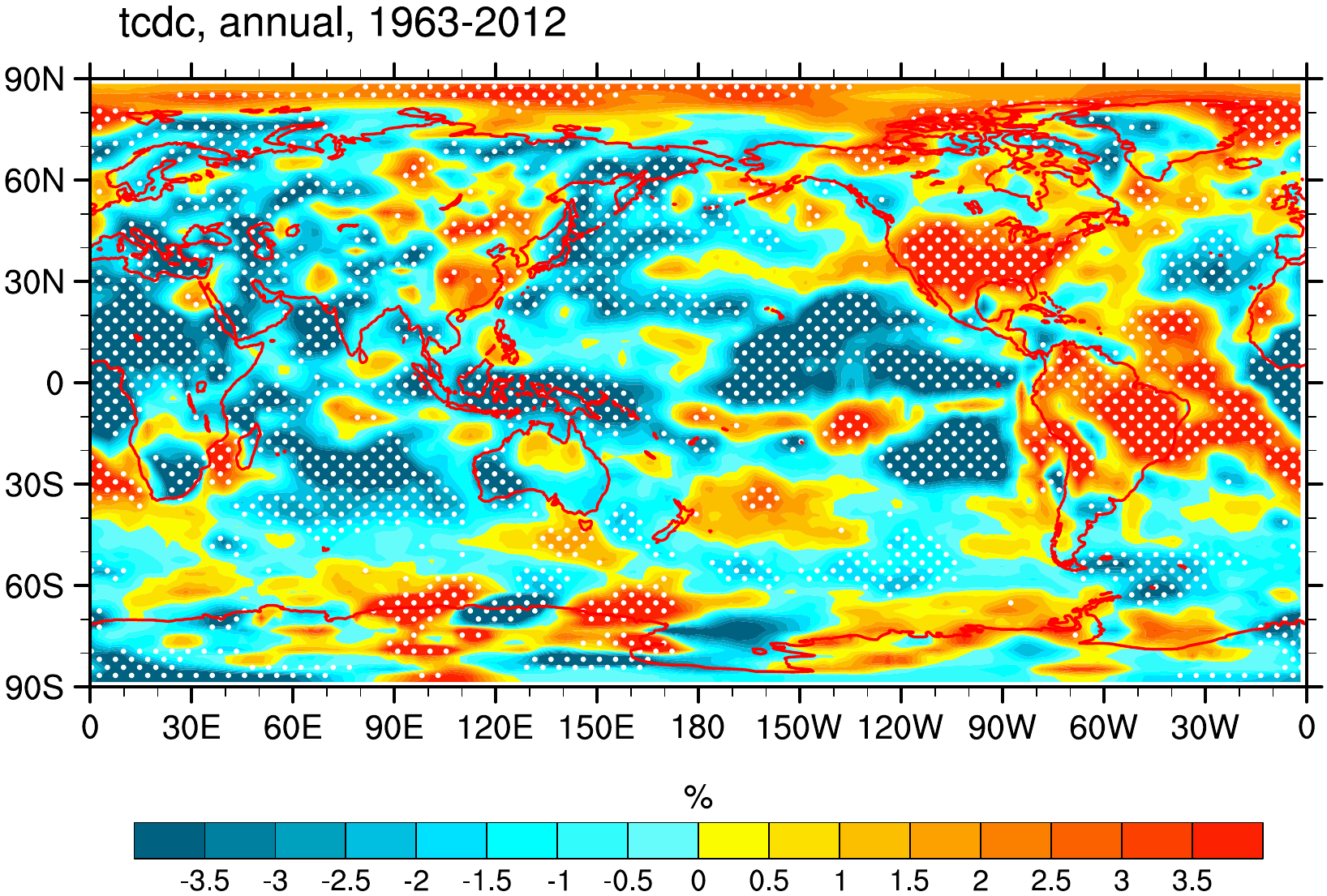}
\noindent\includegraphics[width=13pc]{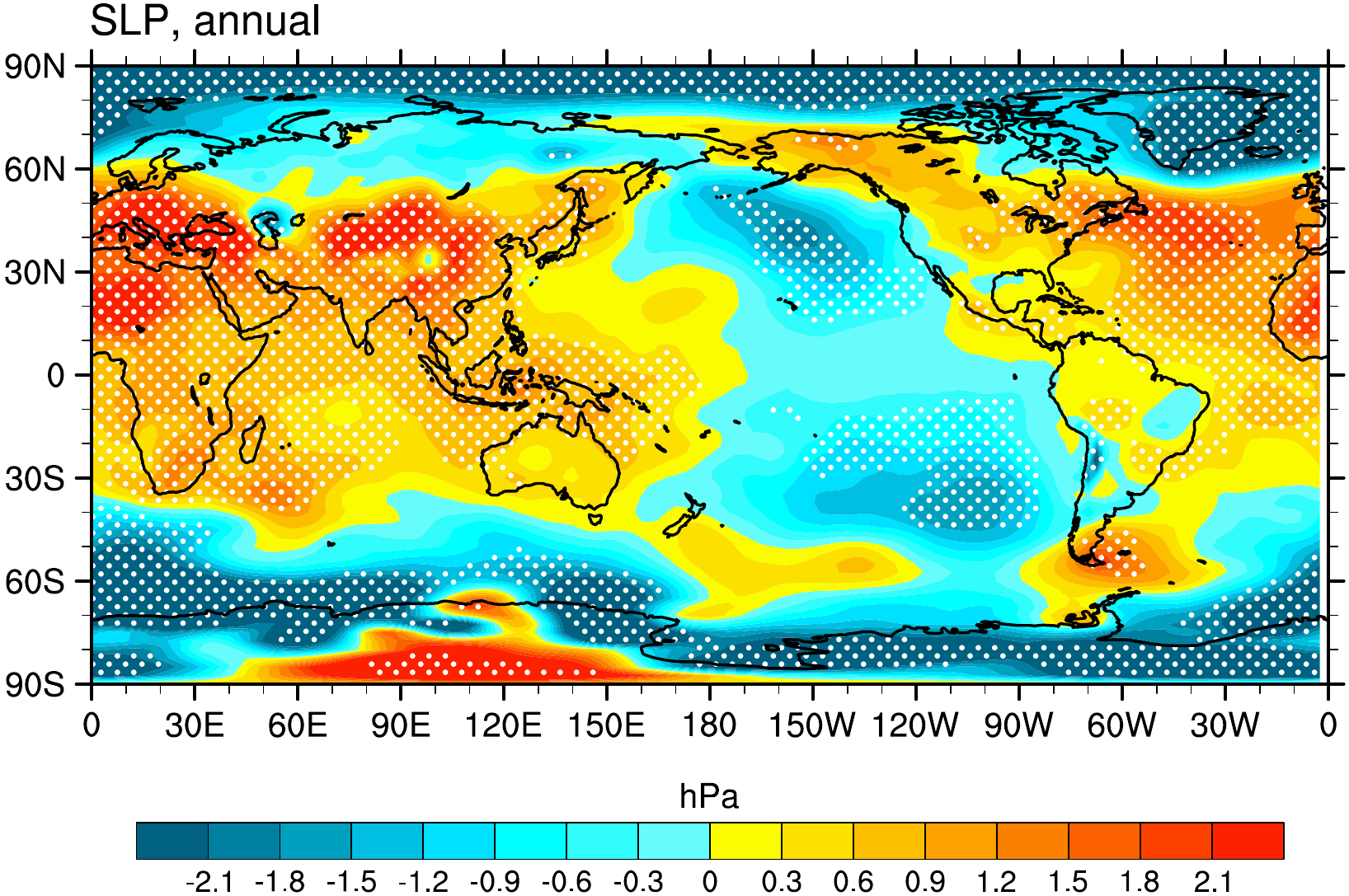}
\noindent\includegraphics[width=13pc]{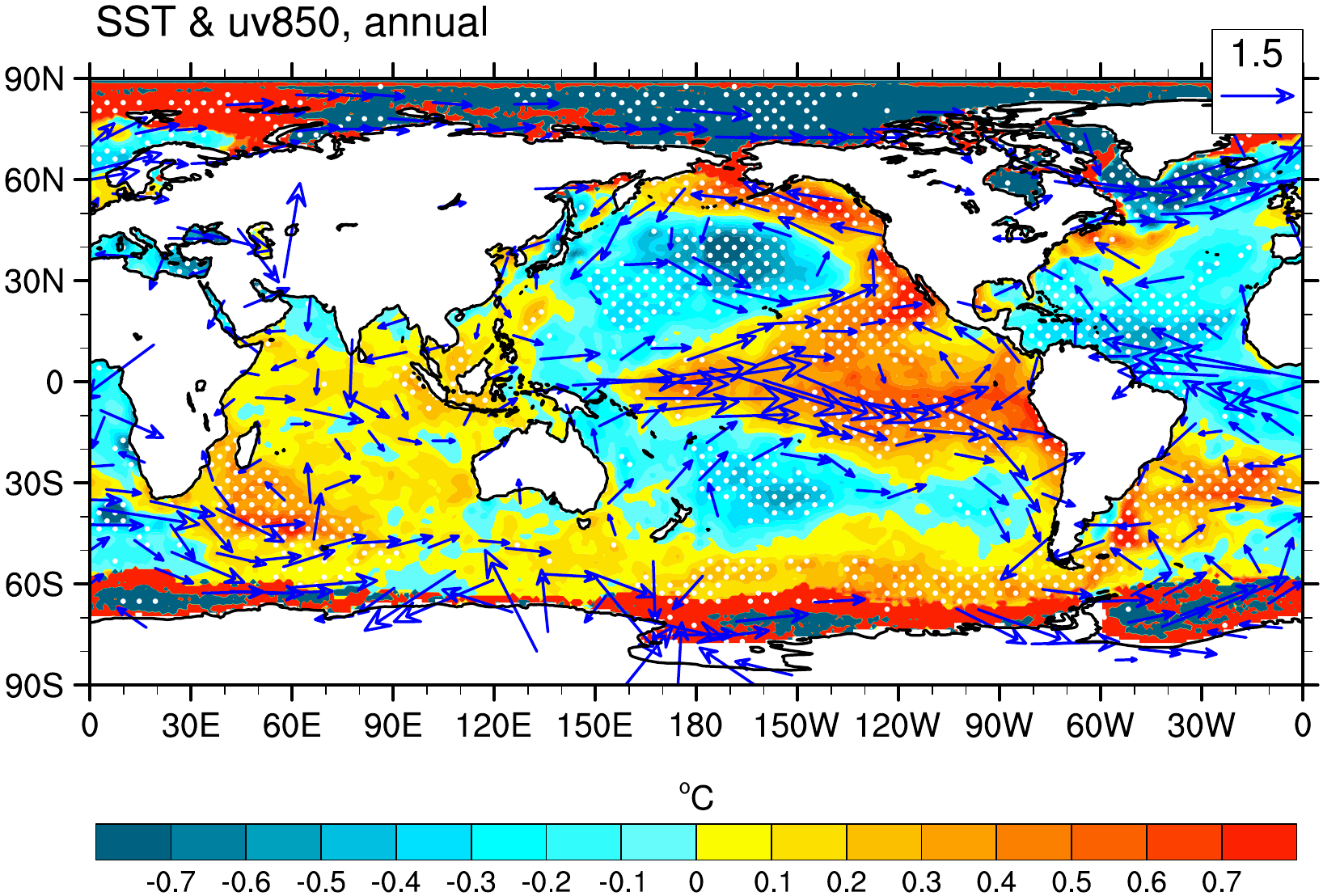}
\noindent\includegraphics[width=13pc]{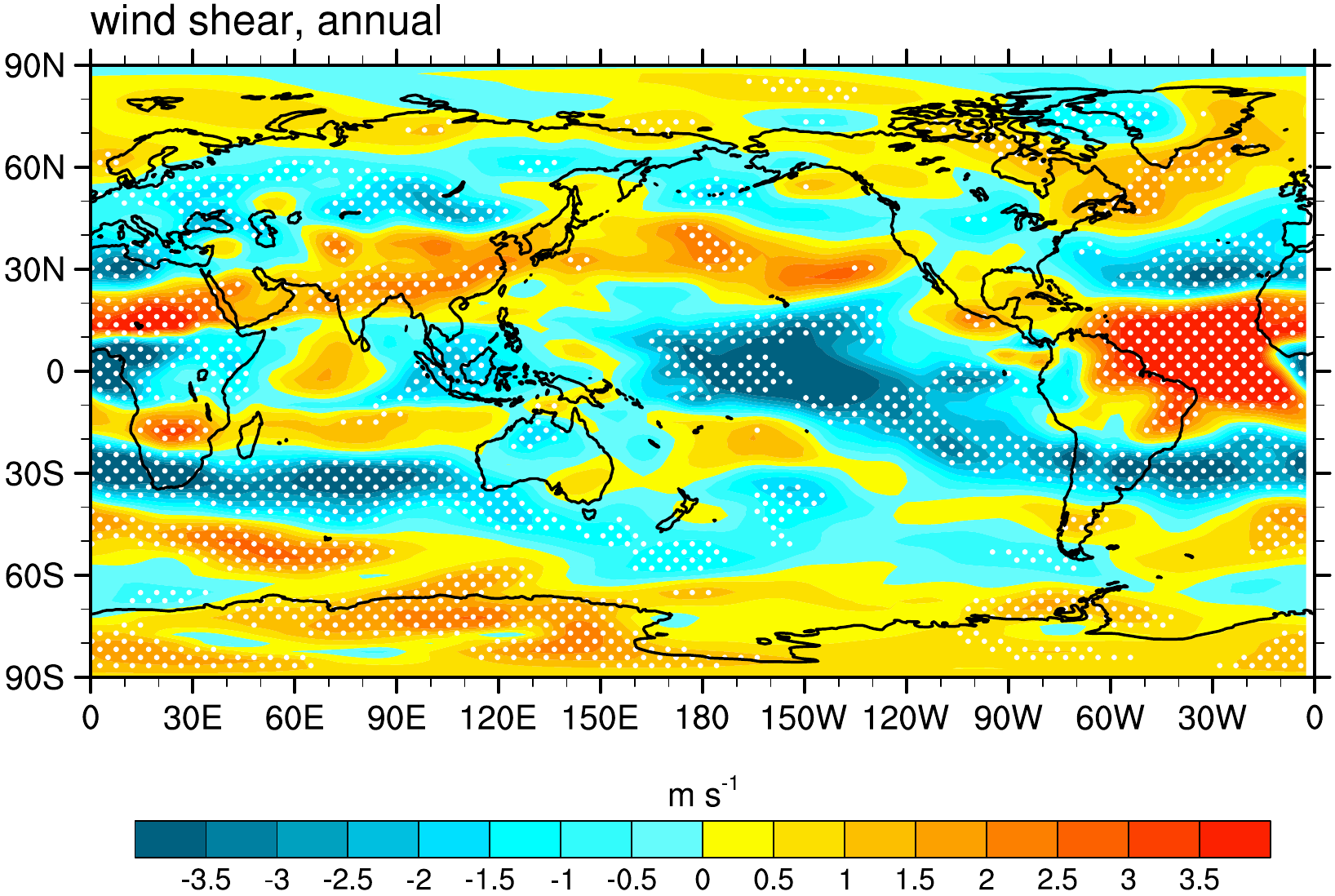}
\caption{Composite maps of some parameters in 1963-2012 between large and small $E_{in}$ years. (A) annual mean total cloud cover (tcdc)). (B) the sea level pressure (SLP). (C) annual mean sea surface temperature (SST)) and 850-hPa wind anomalies ($ms^{-1}$) in blue arrows. (D) the vertical wind shear ($ms^{-1}$). Stippled regions indicate significant correlations at the 90\% confidence level. Large $E_{in}$ years are defined as the standardized $E_{in} \ge$ 1.0 standard deviation, resulting 1982, 1983, 1984, 1989, 1991, 1992, 2003. Small $E_{in}$ years are defined as the standardized $E_{in} \le$ 1.0 standard deviation, resulting 1963, 1964, 1965, 2007, 2008, 2009, 2010, 2011. The map is generated by the the software of NCAR Command Language (Version 6.3.0, 2016, Boulder, Colorado: UCAR/NCAR/CISL/TDD. http://dx.doi.org/10.5065/D6WD3XH5).}
\label{composite}
\end{figure}

Composite maps in Fig. 4 provide indirect evidences to support our interpretation. When the solar wind energy flux increases, 1) the tcdc over tropical Pacific and Indian Oceans decreases, while the situation over tropical NA reverses, as shown in Fig. 4A; 2) the SLP over tropical Pacific Ocean decreases, while it increases over tropical Atlantic and Indian Oceans, as shown in Fig. 4B. It means that the SLP is significantly higher over the western hemisphere and North Atlantic Ocean than that over the eastern Pacific, which would lead to eastward/westward pressure gradient over the tropical Pacific/Atlantic and further influence the zonal wind and SST in situ. As shown in Fig. 4C, the SST over tropical Pacific and Indian Oceans increases, while the situation reverses for tropical Atlantic Ocean. The positive/negative SST anomaly at lower latitudes (30S-30N) is well confined where negative/positive tcdc anomaly is located (Figure 3(A)). This is reasonable due to the existence of cloud-radiation feedback. Meanwhile, the 850-hPa wind anomalies represent west wind enhancement over tropical Pacific and India Oceans, while east wind enhancement over tropical Atlantic Ocean. These features are consistent with the results shown in Fig. 4A and 4B. At last, the vertical wind shear over tropical Pacific Ocean reduces, while the situations reverse for tropical Atlantic and Indian Oceans, as shown in Fig. 4D. The reduction of vertical wind shear and enhancement of SST over tropical Pacific Ocean both contribute to the intensification of tropical cyclone activities in WP, EP, and SP. This is consistent with the positive correlation of tropical cyclone activities at WP, EP, and SP basin on $E_{in}$. The enhancement of vertical wind shear and reduction of SST over tropical Atlantic Ocean contribute to the decrease of tropical cyclone activities in NA, which is consistent with the negative correlation of tropical cyclone activities at NA basin on $E_{in}$. The competitive contributions from enhancements of SST and vertical wind shear over tropical Indian Ocean complicate the variation of tropical cyclone activities at NI and SI basins, and there exists no significant correlations with $E_{in}$ as shown before.

\section{Summary}
\label{sec:sum}
Many studies presented that solar variability do play an significant role in affecting the Earth's climate change. Almost all of previous studies focused on the effects of solar total irradiation energy. As the second major source, the solar wind energy flux exhibits more significant long-term variations, but its effect has been rarely concerned. Although the energy content of solar wind energy flux is of 4-5 orders lower than that of irradiation energy, its long-term variation is much more significant.

For the first time, we find the evidence that the modulation of the solar wind energy flux on the global tropical cyclone activity, and propose a plausible mechanism. We believe this will open a new window to discuss the natural driver of the climate change.  In this study, the global tropical cyclone activity is found to be modulated by solar wind energy flux, but not the solar irradiation and the Earth's weather and climate parameters. A possible mechanism is proposed and some evidences are also presented. The findings are helpful to our understanding of solar impact on the Earth's climate change. More attentions on solar wind energy flux is suggested to be paid in the future studies.

%

\section*{Acknowledgments}
We thank the use of tropical cyclone data from IBTrACS project, the SOI data from CAS/NCAR program, the SST data from The Extended Reconstructed Sea Surface Temperature (ERSST) dataset, and solar wind data from OMNIWeb. We also appreciate NOAA for providing sea surface temperature dataset and NCEP reanalysis datasets including sea level pressure, total cloud cover, and wind, which are available for free at \url{http://www.esrl.noaa.gov/psd/}. This work was supported by 973 program 2012CB825602 and 2012CB957801, NNSFC grants 41574169 and 41204118, and in part by the Specialized Research Fund for State Key Laboratories of China. H. Li is also supported by Youth Innovation Promotion Association of the Chinese Academy of Sciences and NSSC research fund for key development directions.

\section*{References}


\end{document}